\documentclass[twocolumn,aps,prb,epsf]{revtex4}
\usepackage{graphicx}

\begin{document}

\title{Effect of bilayer coupling on tunneling conductance of double-layer high T$_{c}$
cuprates}
\author{Y. H. Su$^{1}$, J. Chang$^{1}$, H. T. Lu$^{2}$, H. G. Luo$^{1}$, and T. Xiang$^{1}$}

\address{$^1$Institute of Theoretical Physics and Interdisciplinary
Center of Theoretical Studies, Chinese Academy of Sciences,
Beijing 100080, China}

\address{$^2$Department of Physics, Peking University, Beijing 100871, China}

\date{\today}

\begin{abstract}
Physical effects of bilayer coupling on the tunneling spectroscopy
of high T$_{c}$ cuprates are investigated. The bilayer coupling
separates the bonding and antibonding bands and leads to a splitting
of the coherence peaks in the tunneling differential conductance.
However, the coherence peak of the bonding band is strongly
suppressed and broadened by the particle-hole asymmetry in the
density of states and finite quasiparticle life-time, and is
difficult to resolve by experiments. This gives a qualitative
account why the bilayer splitting of the coherence peaks was not
clearly observed in tunneling measurements of double-layer
high-T$_c$ oxides.
\end{abstract}

\maketitle

The interlayer coupling of electrons in high T$_c$ cuprates was
predicted to depend strongly on the in-plane momentum and vanish
along the zone diagonals of the 2D Brillouin
zone.\cite{LDA,Xiang1,Xiang2} This would lead to an anisotropic splitting
of the energy bands in bilayer compounds. This bilayer splitting
was first observed by Feng et al in the angle-resolved
photonemission spectroscopy (ARPES) of heavily overdoped (OD)
Bi$_2$Sr$_2$CaCu$_2$O$_{8+x}$ (Bi2212) compounds.\cite{Feng} The
maximum splitting occurs near the anti-nodal points $(\pm \pi ,0)$
and $(0,\pm \pi )$ and varies from $20$ meV in the superconducting
state to $88$ meV in the normal state. Chuang et al also observed
this splitting in OD Bi2212 samples in the normal state, but with
a larger splitting energy $110$ meV.\cite{Chuang} Furthermore, by
analyzing the energy dependence of ARPES spectra, Kordyuk et al
concluded that the peak-dip-hump lineshape observed in ARPES are
stemmed from the bilayer splitting.\cite{Kordyuk}

Tunneling spectroscopy is an important tool for exploring low
energy properties of high T$_c$ superconductors (HTSC), as the
tunneling conductance is proportional to the density of states
(DOS) of electrons. The tunneling measurements reveal important
features of d-wave superconductors, such as the superconducting
coherence peak at the gap edges and the V-shape low-energy
spectrum associated with the linear DOS.\cite{Renner1,DeWilde,
Renner2,Ozyuzer1,Ozyuzer2,Davis1,Davis2} Besides
these d-wave features, an asymmetric tunneling conductance
background with a negative slope has also been observed.\cite
{Renner1,DeWilde,Renner2} In contrast to the ARPES,
the bilayer coupling effect in tunneling experiments has not been reported.

In this paper we investigate the effect of the bilayer coupling on
tunneling measurements. The bilayer coupling splits the energy bands
into the bonding and antibonding ones and leads to a separation of the
superconducting coherence peaks. This property can be used to
probe the bilayer effect from tunneling measurements. Furthermore,
it is shown that the DOS contributed from the bonding and
antibonding bands behave differently in presence of the
particle-hole asymmetry. With a negative slope in the normal
density of states, the particle-hole asymmetry tends to reduce the
DOS of the bonding band, but enhance that of the antibonding band.
This will enhance one coherence peak but reduce the other one, and
eventually cause the disappearance of two-coherence-peak
structure.

Our model involves the anisotropic c-axis coupling between two
CuO$_2$ planes and is defined by the following Hamiltonian
\begin{eqnarray}
H &=&\sum_{l,\mathbf{k}\sigma }\varepsilon
(\mathbf{k})c_{l,\mathbf{k}\sigma }^{\dag }c_{l,\mathbf{k}\sigma
}+\sum_{\mathbf{k}\sigma }t_{\bot }(\mathbf{k})(c_{1,\mathbf{k}\sigma
}^{\dag }c_{2,\mathbf{k}\sigma }+h.c.) \nonumber \\
&&-\sum_{l,\mathbf{k}}\Delta (\mathbf{k})(c_{l, \mathbf{k}\uparrow
}^{\dag }c_{l,\mathbf{-k}\downarrow }^{\dag }
+c_{l,\mathbf{-k}\downarrow }c_{l,\mathbf{k}\uparrow }) ,  \label{eqn2.0}
\end{eqnarray}
where $c_{l,\mathbf{k}\sigma }^{\dag }(l=1,2)$ creates electons in
the $l$'th CuO$_2$ plane with momentum $\mathbf{k}$ and spin
$\sigma $. The kinetic energy $\varepsilon (\mathbf{k})$ includes
the chemical potential and thus the Fermi energy $\varepsilon
_F=0$. The superconducting energy gap is assumed to have
$d_{x^2-y^2}$ symmetry and $\Delta (\mathbf{k})=\Delta _0(\cos
k_x-\cos k_y)/2$. In a tetragonal high T$_c$ cuprate, the c-axis
electron hopping integral is anisotropic:\cite {LDA,Xiang1,Xiang2}
\[
t_{\bot }(\mathbf{k})=-\frac{t_z}4(\cos k_x-\cos k_y)^2.
\]
This anisotropy results from the hybridization between the bonding
O 2p and unoccupied Cu 4s orbitals. If $\Delta
(\mathbf{k})=0$ in one of the double planes, Eq. (\ref{eqn2.0})
is the model that was widely used for studying the proximity
effect in YBCO materials. \cite{Atkinson1995,Xiang1996}

Defining the operators $d_{1,\mathbf{k}\sigma
}=(c_{1,\mathbf{k}\sigma }+c_{2,\mathbf{k}\sigma })/\sqrt{2}$ and
$d_{2,\mathbf{k}\sigma }=(c_{1, \mathbf{k}\sigma
}-c_{2,\mathbf{k}\sigma })/\sqrt{2}$, we can decouple the above
Hamiltonian into two independent parts,
\begin{eqnarray}
H &=&H_1+H_2, \\
H_i &=&\sum_{\mathbf{k}\sigma }\varepsilon
_i(\mathbf{k})d_{i,\mathbf{k} \sigma }^{\dag
}d_{i,\mathbf{k}\sigma } \nonumber \\
&&-\sum_{\mathbf{k}}\Delta (\mathbf{k}
)\left( d_{i,\mathbf{k}\uparrow }^{\dag
}d_{i,\mathbf{-k}\downarrow }^{\dag }+h.c.\right) , \label{eqn2.1.2}
\end{eqnarray}
with $\varepsilon _{1,2}(\mathbf{k})=\varepsilon (\mathbf{k})\pm
t_{\bot }( \mathbf{k})$. $H_{1,2}$ are the BCS Hamiltonians for
the bonding and antibonding bands, respectively. From the
decoupled Hamiltonian, we can readily obtain the energy spectra of
the Bogoliubov quasiparticles,
\[
H_i=\sum_{\mathbf{k}}E_i(\mathbf{k})\left( \gamma _{i,\mathbf{k}\uparrow
}^{\dag }\gamma _{i,\mathbf{k}\uparrow }+\gamma _{i,\mathbf{-k}\downarrow
}^{\dag }\gamma _{i,\mathbf{-k}\downarrow }\right) ,
\]
where $E_{1,2}(\mathbf{k})=\sqrt{\Delta
^2(\mathbf{k})+(\varepsilon (\mathbf{ k})\pm t_{\bot
}(\mathbf{k}))^2}$ and the Bogoliubov quasiparticle operators are
defined by
\begin{eqnarray}
\gamma _{i,\mathbf{k}\uparrow } &=&u_{i,\mathbf{k}}d_{i,\mathbf{k}\uparrow
}-v_{i,\mathbf{k}}d_{i,\mathbf{-k}\downarrow }^{\dag },  \label{eqn2.1.4} \\
\gamma _{i,\mathbf{-k}\downarrow }^{\dag }
&=&v_{i,\mathbf{k}}d_{i,\mathbf{k} \uparrow
}+u_{i,\mathbf{k}}d_{i,\mathbf{-k}\downarrow }^{\dag },
\label{eqn2.1.5}
\end{eqnarray}
and
\begin{eqnarray}
u_{i,\mathbf{k}}^2 &=&\frac 12\left( 1+\frac{\varepsilon
_i(\mathbf{k})}{E_i(
\mathbf{k})}\right) ,  \label{eqn2.1.6} \\
v_{i,\mathbf{k}}^2 &=&\frac 12\left( 1-\frac{\varepsilon
_i(\mathbf{k})}{E_i( \mathbf{k})}\right) .  \label{eqn2.1.7}
\end{eqnarray}

The energy spectra of the two quasiparticle bands demonstrate the
bilayer splitting as observed in ARPES.\cite{Feng, Chuang} Because
of the anisotropy of $t_{\bot }(\mathbf{k})$, the largest
splitting occurs at the Fermi surface in the vicinity of $(\pm \pi
,0),(0,\pm \pi )$.

The density of states of electrons $\rho (\omega )$ is defined by
the imaginary part of the retarded Green function $G_{i,\sigma
}^R(\mathbf{k} ,\omega )$ of fermions
$\{d^{\dagger}_{i,\mathbf{k}\sigma}, d_{i,\mathbf{k}\sigma}\}$,
\begin{equation}
\rho (\omega )=-\frac 1{\pi N}\sum_{i\mathbf{k}\sigma }\text{Im}G_{i,\sigma
}^R(\mathbf{k},\omega ),
\label{eqna.1}
\end{equation}
where $N$ is the total number of $\mathbf{k}$ vectors in the first
Brillouin zone. For the $i$'th band, the density of states is
given by $\smallskip $
\[
\rho _i(\omega )=-\frac 2{N\pi }\sum_{\mathbf{k}}\text{Im}\left(
\frac{u_{i, \mathbf{k}}^2}{\omega -E_i(\mathbf{k})+i\Gamma}
+\frac{v_{i,\mathbf{k}}^2}{ \omega +E_i(\mathbf{k})+i\Gamma}\right) .
\]
$\Gamma$ is the quasiparticle scattering rate which origins
from the lifetime effects, stoichiometry variations, noise
smearing, etc.\cite{Dynes1978,Cucolo1992} It can be also
taken as a free parameter associated with the energy resolution in
the tunneling experiment if the scattering rate is smaller than
the experimental resolution.

Since the low-energy physics is governed by excitations near the
Fermi energy, we assume that the kinetic energy depends only on
the absolute value of the momentum, i.e., $\varepsilon
(\mathbf{k})=\varepsilon (k)$. The anisotropic d-wave gap function
and c-axis coupling $t_{\bot }(\mathbf{k})$ can also be simplified
as $\Delta _i(\mathbf{k})=\Delta _{i,0}\cos(2\phi )$ and $t_{\bot
}(\mathbf{k})=-t_z\cos^2(2\phi )$ in the vicinity of the Fermi
surface.

Near the Fermi energy, the normal density of states can be written
as
\begin{equation}
\rho _N(\varepsilon )\simeq \rho _N(0)+\rho _N^{\prime
}(0)\varepsilon
\label{eqna.3}
\end{equation}
up to the leading order approximation in $\varepsilon$, where
$\rho _N(0)$ is the normal DOS at the Fermi energy. $\rho
_N^{\prime }(0)$ is the linear coefficient of the DOS. $\rho
_N^{\prime }(0)$ is finite if particle-hole symmetry is broken. A
number of ARPES experiments have shown that there is a flat band
at about $200$ meV below the Fermi energy in deeply underdoped
cuprates.\cite{Saddle1,Damascelli2003}
The presence of this flat band is an indication of Van
Hove singularity and suggests that the variation of the DOS around
the Fermi surface can no longer be neglected as in conventional
metals. Moreover, as revealed by the tunneling measurements, the
tunneling conductance varies almost linearly with the applied bias
around the Fermi energy in the normal state.\cite{Renner1,DeWilde,Renner2}
It suggests that particle-hole symmetry is broken in high-Tc cuprates. Thus, it is
important to include the particle-hole asymmetric term in the
analysis of tunneling measurement data. The linear approximation
of DOS, defined by Eq. (\ref{eqna.3}), is valid if
the energy of the Van Hove singularity is close to the Fermi
energy but still much lower than
the energy range we are interested in.

With the above equations, it is straightforward to show that
\begin{eqnarray}
\rho _i(\omega )&\simeq& \rho _N(0)I_{i,1}(\omega )+\rho _N^{\prime
}(0)sgn(\omega )I_{i,2}(\omega ) \nonumber \\
&&\pm t_z\rho _N^{\prime }(0)I_{i,3}(\omega ),
\label{eqn2.4.1}
\end{eqnarray}
where $I_{i,l}(\omega )$ are
\begin{eqnarray}
I_{i,1}(\omega ) &=&\frac 1{2\pi ^2}\int d\phi \int d\varepsilon \frac
\Gamma {\left( \left| \omega \right| -\Omega_{i}\right) ^2+\Gamma ^2},
\label{eqn2.4.2} \\
I_{i,2}(\omega ) &=&\frac 1{2\pi ^2}\int d\phi \int d\varepsilon
\frac{ \varepsilon ^2}{\Omega_{i}}\frac \Gamma {\left( \left|
\omega \right|-\Omega_{i}\right) ^2+\Gamma ^2},  \label{eqn2.4.3} \\
I_{i,3}(\omega ) &=&\frac 1{2\pi ^2}\int d\phi \int d\varepsilon \frac{\cos
^2(2\phi )\Gamma }{\left( \left| \omega \right| -\Omega_{i}\right)
^2+\Gamma ^2},  \label{eqn2.4.4}
\end{eqnarray}
and $\Omega_{i}\equiv\sqrt{\Delta _{i,0}^2\cos ^2(2\phi
)+\varepsilon ^2}$ .
The first term at the right-hand side of Eq.(\ref{eqn2.4.1}) has the largest
contribution to the DOS. The supperconducting coherence peaks are located at
the gap edge, namely at $\omega _i=\pm \Delta _{i,0}$. In the limit
$\Gamma\rightarrow 0^{+}$, $I_{i,1}(\omega )\propto \left| \omega \right|$
near the Fermi energy, thus the low
energy DOS of quasiparticles is linear $\rho _i(\omega )\simeq \rho
_N(0)\left| \omega \right| /\Delta _{i,0}$. These special d-wave characters
have already been observed in the tunneling measurements\cite
{Renner1,DeWilde,Renner2,Ozyuzer1,Ozyuzer2,Davis1,Davis2}.

Particle-hole symmetry is broken by the second term in
(\ref{eqn2.4.1}). The asymmetric DOS induced by this term has been observed
in STM or other tunneling spectra. \cite{Renner1,DeWilde,Renner2}
Since $I_{i,2}(\omega )$ is positive, a negative $\rho _N^{\prime }(0)$
will enhance the DOS below the Fermi energy, but reduce that above
the Fermi energy.

The bilayer coupling appears in the third term of Eq.
(\ref{eqn2.4.1}). It leads to the difference in the DOS of the
bonding and antibonding bands. This difference is proportional to
both the c-axis hopping integral $t_z$ and $\rho _N^{\prime }(0)$.
For a system with $\rho _N^{\prime }(0)<0$, the bonding band DOS
$\rho _1(\omega )$ is reduced and the antibonding band DOS $\rho
_2(\omega )$ is enhanced.

\begin{figure}
\begin{center}
\includegraphics[width=8cm]{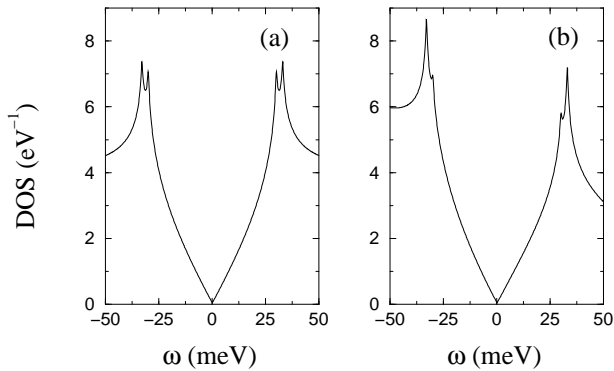}
\renewcommand{\figurename}{Fig}
\caption{Density of states in a bilayer compound. The parameters
used are $\Delta_{1,0}=30$ meV, $\Delta_{2,0}=33$  meV,
$t_{z}=50$ meV, $\Gamma=0.25$ meV, $\rho_{N}(0)=1$ eV$^{-1}$, (a)
$\rho^{\prime}_{N}(0)=0$, (b) $\rho^{\prime}_{N}(0)=-8$ eV$^{-2}$.}
\end{center}
\end{figure}

\begin{figure}
\begin{center}
\includegraphics[width=8cm]{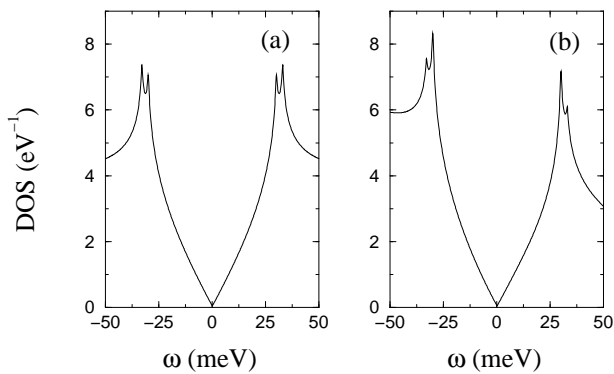}
\renewcommand{\figurename}{Fig}
\caption{Same as for Fig. 1 but with $\Delta_{1,0}=33 $ meV,
$\Delta_{2,0}=30$ meV.}
\end{center}
\end{figure}

Because of the c-axis coupling, the coherence peaks of the bonding
and antibonding bands are separated. The maxima of the energy gap
for both bonding and antibonding bands are located near the four
points $(\pm \pi ,0),(0,\pm \pi )$. Therefore, the largest
difference in the energy gap, $\delta _s\equiv \left| \Delta
_{1,0}-\Delta _{2,0}\right| $, also occurs near these positions,
where $\Delta _{i,0}\simeq \Delta(\mathbf{k} _{i,F})$ is the gap
maximum of the i'th band at the Fermi momentum ${\bf k}_i$ along
the anti-nodal direction. From the ARPES data in Ref.[4], we
estimate the difference between the anti-nodal Fermi momentum to
be $|{\bf k}_{1,F}-{\bf k}_{2,F}| \simeq 0.12\pi$ and the difference of
the largest energy gaps to be $\delta _s\simeq 0.1\Delta _0$. In
OD Bi2212 compounds, $\Delta _0\simeq 30$ meV, thus $\delta
_s\simeq 3$ meV.

Fig. 1 shows the DOS for a system with $\Delta _{1,0}=30$ meV,
$\Delta _{2,0}=33$ meV, and $\rho _N(0)=1$ eV$^{-1}$. The results for both
$\rho _N^{\prime }(0)=0$ and $\rho _N^{\prime }(0)=-8$ eV$^{-2}$ are shown.
In the case $\rho _N^{\prime }(0)=0$, two coherence peaks appear
due to the bilayer splitting. These two peaks, at $30$ meV and
$33$ meV, come from the bonding and antibonding bands,
respectively. When $\rho _N^{\prime }(0)=-8$ eV$^{-2}$, the DOS becomes
non-symmetric, as expected. This asymmetric feature exists in both
the normal and superconducting states, consistent with the
reported data.\cite{Davis1,Davis2} However, the peak of the bonding band
is strongly suppressed.

For comparison, Fig. 2 shows the DOS for a system with $\Delta _{1,0}=33$ meV
and $\Delta _{2,0}=30$ meV. The asymmetric behavior shown in Fig. 2
(b) is similar to the case for Fig. 1 (b), but the higher
coherence peak is located at lower energy.

\begin{figure}
\begin{center}
\includegraphics[width=8cm]{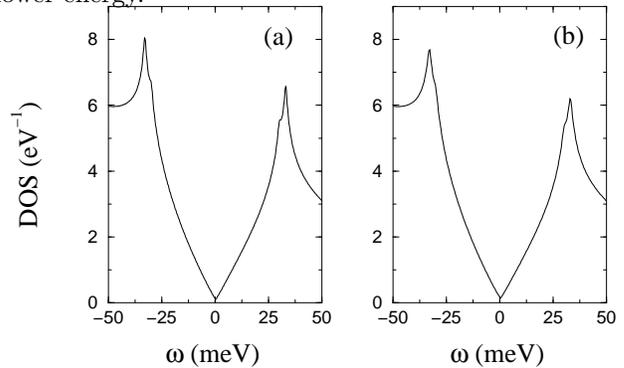}
\renewcommand{\figurename}{Fig}
\caption{Density of states with   (a) $\Gamma=0.5$  meV  (b)
$\Gamma=0.75$ meV. Other parameters used are the same as for Fig.
1 (b)}
\end{center}
\end{figure}

Compared with the case $\rho _N^{\prime }(0)=0$, the particle-hole
asymmetry suppresses strongly the lower coherence peak. If the
quasiparticle scattering rate $\Gamma$ is large or the energy
resolution is not high enough, it is certainly difficult to
resolve this double-coherence-peak structure in the tunneling
spectra. This is explicitely illustrated in Fig. 3.
With increasing $\Gamma $, the double-peak structure
disappears gradually and the lower coherence peak becomes
indistinguishable from the background. Therefore, in order to
observe this bilayer splitting in tunneling spectra, experimental
measurements with high quality single crystal and high energy
resolution are desired.

In conclusion, the superconducting coherence peaks are separated
in bilayer high-T$_c$ superconductors and can be used to probe the
bilayer coupling effect with tunneling measurements. The
particle-hole asymmetry in the DOS enhances one of the coherence
peaks, but reduces another one. If the life-time of quasiparticles
is very short or the energy resolution in tunneling measurements
is not high enough, the lower coherence peak is difficult to be
resolved from the conductance background. This gives a qualitative
account why the bilayer splitting has not been unambiguously
observed in tunneling spectra.

This work was supported by the National Natural Science Foundation of
China and the special funds for Major State Basic Research Projects of
China.

\end{document}